\documentclass[twocolumn,prl]{revtex4}
\usepackage{graphics,graphicx,epsfig}
\usepackage{epsf,epstopdf,wrapfig}
\usepackage{amssymb,amsfonts,amsmath}
\usepackage{graphicx,pstricks,epsfig}
\include{epsf}

\newcommand{\degree}{{}^{\rm o}}

\newcommand{\proton}{{\rm H}^{+}}
\newcommand{\cf}{{\em cf.~}}
\newcommand{\ie}{{\em i.e.~}}

\begin{document}
\title{Modeling torque versus speed, shot noise, and rotational diffusion of the bacterial flagellar motor}

\author{Thierry Mora${}^{1,2}$, Howard Yu${}^{2}$ and Ned S. Wingreen${}^{3}$}

\address{${}^{1}$Lewis-Sigler Institute for Integrative Genomics, ${}^{2}$Joseph Henry Laboratories of Physics, ${}^{3}$Department of Molecular Biology,
Princeton University, Princeton, New Jersey, USA}


\begin{abstract}
We present a minimal physical model for the flagellar motor that enables bacteria to swim. Our model explains the experimentally measured torque-speed relationship of the proton-driven {\em E. coli} motor at various pH and temperature conditions. In particular, the dramatic drop of torque at high rotation speeds (the ``knee'') is shown to arise from saturation of the proton flux. Moreover, we show that shot noise in the proton current dominates the diffusion of motor rotation at low loads. This suggests a new way to probe the discreteness of the energy source, analogous to measurements of charge quantization in superconducting tunnel junctions.
\end{abstract}

\maketitle

\begin{figure}
\includegraphics[width=\linewidth]{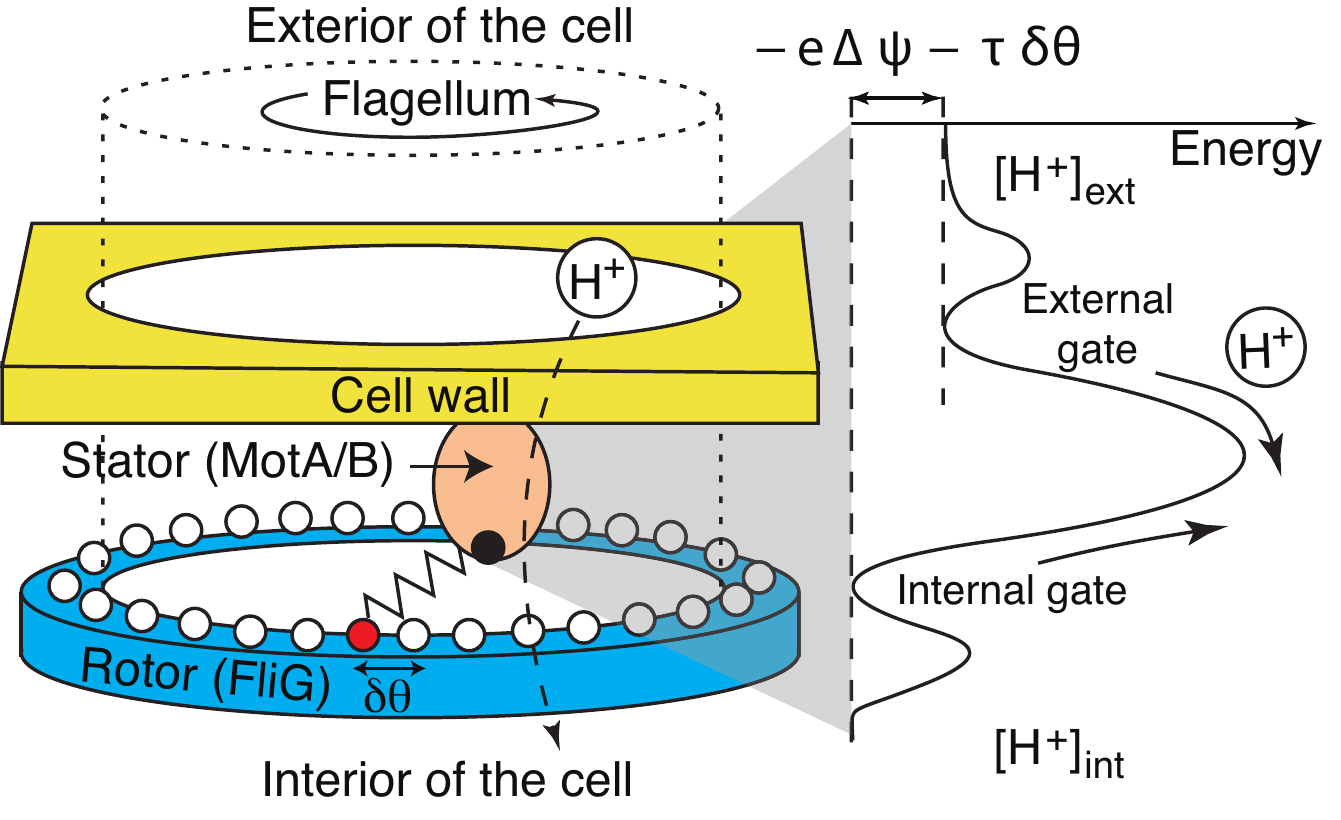}
\caption{Schematic model of the bacterial flagellar motor. Left: The passage of a proton, or possibly a group of protons, through a torque generating unit (a MotA/B stator---only one stator of about $10$ is shown) causes a protein spring to stretch to its next attachment site, represented by circles, on the rotor. Right: To translocate a proton must pass through an external gate, over a barrier, and finally through an internal gate, with all the steps assumed to be reversible. The net energy difference driving proton translocation is the electrical potential energy, $-e\Delta \psi$, minus the work, $\tau\,\delta\theta$, necessary to stretch the protein spring by $\delta\theta$, where $\tau$ is the torque applied by the spring to the rotor.
\label{fig:model}
}
\end{figure}

The bacterial flagellar motor is a
molecular machine that rotates a helical filament and thereby powers the swimming of bacteria like {\em E. coli} \cite{Sowa:2008p2}. Motor rotation is typically driven by $\proton$ ions that generate torque by passing into the cell via the motor, down an electro-chemical gradient called the proton motive force (PMF). Although much work has been devoted to understanding proton translocation and its coupling to torque generation, biochemical details are lacking and many questions remain unanswered. An important one is whether ion translocation is cooperative, \ie whether protons translocate individually or in groups. Here, we present a minimal physical model for torque generation (Fig.~\ref{fig:model}) that not only explains a variety of previous experimental observations, but also suggests a way to measure the cooperativity of proton translocation. Specifically, the model predicts that at low loads, motor diffusion is dominated by proton shot noise with a strong
(quadratic) dependence on proton cooperativity. 

\begin{figure}
\includegraphics[width=\linewidth]{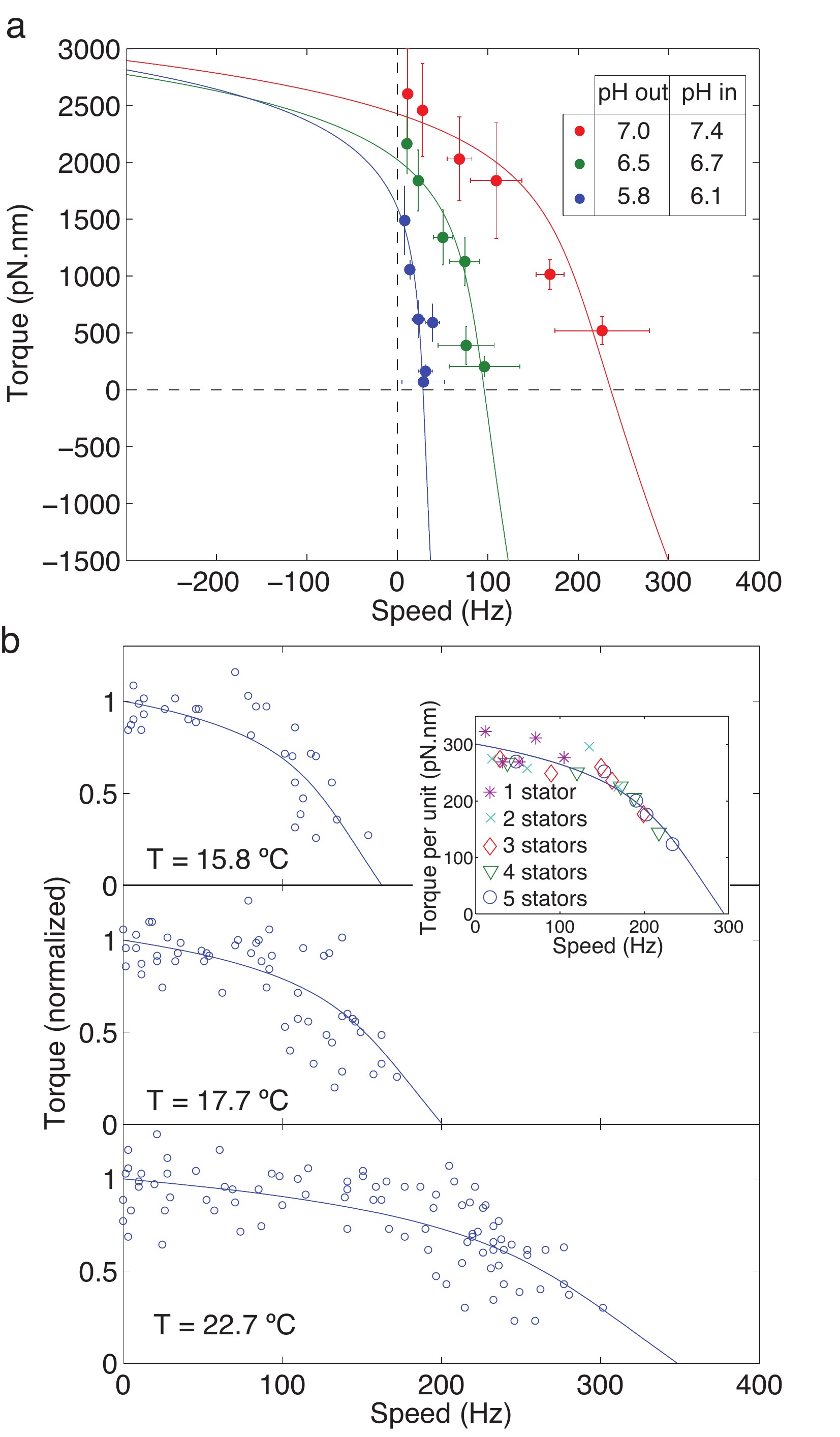}
\caption{The torque-speed relationship {(TSR)} of the {\em E. coli} flagellar motor. a. Rotation speed is measured with beads of various loads (data points) attached to a flagellar stub, under different pH conditions (colors) \cite{Nakamura:2009p9}; solid curves: model fits. b. Rotation speed measured with 0.25--0.36 $\mu$m diameter beads attached to flagellar stubs for a wide range of viscosities, at three different temperatures \cite{Chen:2000p279}; circles: experimental data, solid curves: model fits. Inset: Total torque {\em vs.} speed from \cite{Ryu:2000p204}, normalized by the number of stators. Data collapse indicates that stators contribute independently and additively to the total torque. Solid curve: model {TSR} for a single stator using the same parameters as in the main figure, with the temperature fit as
 $21\degree$C and the stall torque fit as $300$ pN.nm. 
\label{fig:fitnakamura}
}
\end{figure}

The flagellar motor operates with near-perfect efficiency at low speeds \cite{Meister:1987p1090}. As the speed is increased, e.g. by reducing the load, the torque and efficiency initially remain high---the ``plateau'' of the torque-speed relationship {(TSR)}---and then drop abruptly at a  ``knee''  (\cf Fig.~\ref{fig:fitnakamura}). This knee occurs at higher speeds as temperature is increased.
Despite much experimental \cite{Berg:1993p329,Chen:2000p279,Ryu:2000p204,Nakamura:2009p9} and modeling progress \cite{Meister:1989p233,Iwazawa:1993p367,Berry:1999p403,Ryu:2000p204,Xing:2006p99,Meacci:2009p5}, the origin of the knee is still poorly understood. In \cite{Xing:2006p99,Bai:2009p1409} the cause of the knee was argued to be the gating of proton translocation { by the relative position between stator and rotor}. In \cite{Meacci:2009p5}, a detailed model of motor kinetics was proposed { to explain the observation that motor speed is independent of the number of stators at low loads \cite{Yuan:2008p48}}.
In our model, proton translocation, which is assumed to be the rate-limiting step, is modeled by a barrier crossing event.
The knee in the {TSR} then arises from the kinetically limited rate of proton translocation.  {Importantly, our model fully incorporates proton thermodynamics and yields the separate dependence of the {TSR} on the electrical and chemical parts of the PMF.}

Three ingredients underlie our model: ({\em i}) Each torque-generating unit (MotA/B stator) contributes independently and additively to the total torque, in agreement with experimental observations (Fig.~\ref{fig:fitnakamura}b) \cite{Yuan:2008p48}. ({\em ii}) The torque from each MotA/B stator is applied to the rotor by a protein spring.
Proton translocation into the cell causes the stretching of a protein spring to its next attachment site (Fig.~\ref{fig:model}) \cite{Meister:1989p233,Mora:2009p1088}. This assumption enforces the tight coupling between proton current and rotation speed
\cite{Meister:1987p1090}.
({\em iii}) Assuming a cooperativity index of $n$, translocation occurs through three reversible steps: first $n$ protons load into an external gate, then all $n$ cross an energy barrier to an internal gate, and finally all $n$ are released into the cell. The barrier crossing event is the rate-limiting step.
The external and internal gates are necessary to explain the non-linear dependence of the {TSR} on proton concentrations (Fig.~\ref{fig:fitnakamura}a). (We define the distance between two attachment sites as $n\delta\theta$, so that the average displacement per proton is $\delta\theta$.)

The external and internal gates are assumed to be in fast equilibrium with the external and internal proton concentrations, respectively. Their dissociation constants are denoted by $K_{\rm ext}$ and $K_{\rm int}$, so that the occupancy of the external gate is $H_{\rm ext}^n/(K_{\rm ext}^n+H_{\rm ext}^n)$, where $H_{\rm ext}$ is the proton concentration outside the cell, and similarly for the internal gate.
The energy difference between the internal and external gates is $n\times\left[e\Delta\psi+\tau\,\delta\theta+k_BT\log(K_{\rm int}/K_{\rm ext})\right]$. $\Delta\psi$ is the transmembrane electric potential ($\psi_{\rm int}-\psi_{\rm ext}$) and $\tau\,\delta\theta$ is the work necessary to stretch the protein spring.
When $\Delta\psi=0$ and $\tau=0$, the energy barrier per proton to inward translocation is $U_{\rm in}^0$, and the barrier to outward translocation is $U_{\rm out}^0$, with $U_{\rm in}^0-U_{\rm out}^0=k_BT\log(K_{\rm int}/K_{\rm ext})$. In general, some fraction $\alpha$ of the electric potential and some fraction $\beta$ of the work contribute to the inward barrier, so that the barriers per proton to inward and outward translocations are, respectively, $U_{\rm in}=U_{\rm in}^0+\alpha\,e\Delta\psi+\beta \,\tau\,\delta\theta$ and $U_{\rm out}=U_{\rm out}^0-(1-\alpha)e\Delta\psi-(1-\beta)\tau\delta\theta$. For simplicity we have neglected any dependence of $K_{\rm int,ext}$ on $\Delta\psi$ and $\tau$.
Then the rate of inward proton translocations is:
\begin{equation}\label{eq:Jincoop}
J_{\rm in}=nJ_0\frac{H_{\rm ext}^n}{K_{\rm ext}^n+H_{\rm ext}^n}
\frac{K_{\rm int}^n}{K_{\rm int}^n+H_{\rm int}^n}
\exp\left(-n\frac{U_{\rm in}}{k_BT}\right),
\end{equation}
$J_0$ is a kinetic constant (in Hz), and the other prefactors represent the occupancies of the external and internal gates. The outward rate $J_{\rm out}$ is given by a similar expression, so that the net inward proton flux is:
\begin{equation}\label{eq:Jtot}
\begin{split}
&J_{\rm in}-J_{\rm out}=nJ_0 e^{-nU^0_{\rm in}/k_BT}  \frac{H_{\rm ext}^n}{K_{\rm ext}^n+H_{\rm ext}^n}
\frac{K_{\rm int}^n}{K_{\rm int}^n+H_{\rm int}^n}
\\
&\times e^{-n{(\alpha\Delta\psi+\beta\tau\delta\theta)/{k_BT}}}
\left[1-\exp\left(n\frac{e\Delta p+\tau\delta\theta}{k_BT}\right)\right].
\end{split}
\end{equation}
$\Delta p:=\Delta \psi+(k_BT/e)\log(H_{\rm int}/H_{\rm ext})$ is the PMF composed of the electrical and chemical potential differences. Its value is approximately  $-150$ mV in normal conditions.
To account for the data, we assume that the height of the barrier may depend on temperature, and we expand the prefactor to linear order in temperature:
$J_0 e^{-U^0_{\rm in}(T)/k_BT}=\tilde J_0 e^{\eta (T-T_0)}$, where $T_0=17.7\degree$C, from Fig.~\ref{fig:fitnakamura}b, is chosen as a reference temperature.

Rotation is then described by coupled stochastic equations for the angular position of the rotor $\theta$ and the stretching of the protein springs $i=1,\ldots,N$, where $N$ is the number of stators, each exerting a torque $\tau_i$ on the rotor:
\begin{eqnarray}
\frac{d\theta}{dt}&=&\frac{1}{\nu}\sum_{i=1}^N \tau_i+\xi(t),\label{eq:1}\\
\frac{d\tau_i}{dt}&=&k(\tau_i)\left[-\frac{d\theta}{dt}+\Omega(\tau_i)+\xi^{\tau}_i(t)\right].
\label{eq:2}
\end{eqnarray}
$\nu$ is the frictional drag coefficient of the load, and $k(\tau)$ is the spring constant of the (possibly non-Hookean) protein springs. 
The spring constant need not be specified as none of the observables computed below depend on it.
In the second equation, each spring relaxes as the rotor moves ({back-reaction of the rotor onto stators, }$-d\theta/dt$), but gets restreched by proton translocations ($\Omega(\tau_i):=(J_{\rm in}-J_{\rm out})\delta\theta$).
$\xi(t)$ is the thermal noise on the load and satisfies the Einstein relation: $\langle \xi(t)\xi(t')\rangle=2(k_BT/\nu) \delta(t-t')$. $\xi^{\tau}_i$ is the shot noise at each stator $i$, due to the randomness of proton translocation events. To obtain the average speed $\omega$ and torque per stator $\tau=\langle\tau_i\rangle$, we solve Eqs.~(\ref{eq:1},\ref{eq:2}) at steady state ($d\tau_i/dt=0$) and in the absence of noise, yielding: $\omega= d\theta/dt=N\tau /\nu$ and $\tau_i=\tau$, with $\Omega(\tau)=\omega$. Along with Eq.~\eqref{eq:Jtot}, this gives a closed set of equations from which we obtain the {TSR}. { Note that the resulting expression depends separately on the electrical and chemical potential for protons.}
At stall ($\nu\to\infty$) the system is in equilibrium. The energy necessary for a protein spring to move to its next attachment site, which is proportional to the torque $\tau$ it exerts on the rotor, is matched by the PMF, $\tau\delta\theta+e\Delta p=0$.
Consequently the total torque grows linearly with the PMF, $N\tau=-Ne\Delta p/\delta \theta$, and the efficiency near stall is $\approx 100\%$, in agreement with experiments \cite{Fung:1995p1109,Meister:1987p1096}.

Our model with no cooperativity ($n=1$) can fit all existing measured {TSR} of the {\em E. coli} motor. Some our model's parameters are fixed properties of the motor and thus are fit by single values: ($\delta\theta=4.6\degree$, $K_{\rm int}=1.2\cdot 10^{-8}$, $K_{\rm ext}=2\cdot 10^{-7}$, $\tilde J_0=670$ Hz, $\alpha=0.2$, $\beta=0.078$,  $\eta=0.11\,{\rm K}^{-1}$) while others depend on conditions ($T$, $H_{\rm int,ext}$, $\Delta \psi$, number of stators $N$), and may or may not have been measured in the experiments.

\begin{figure}
\includegraphics[width=\linewidth]{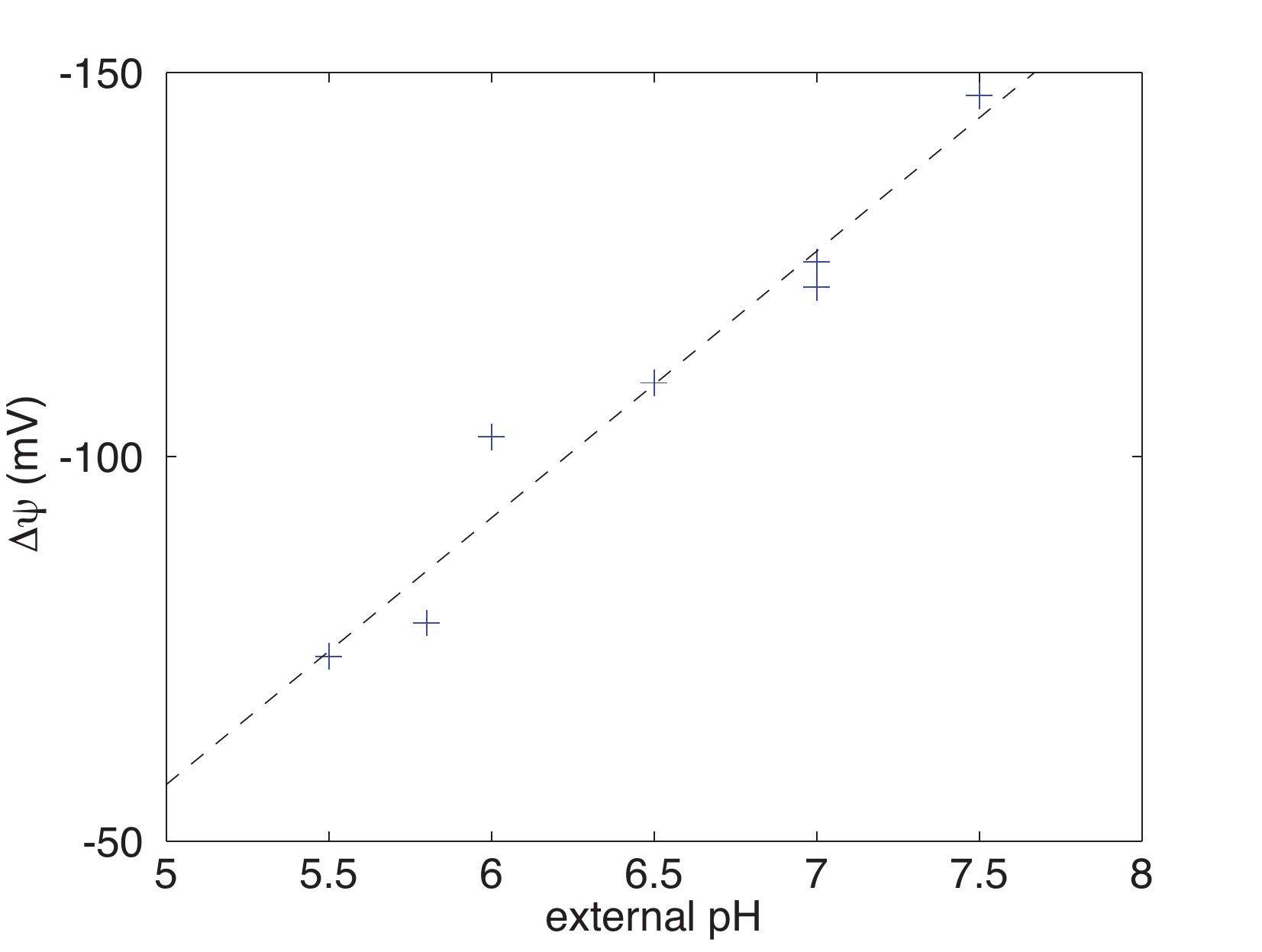}
\caption{Electric potential difference versus external pH, as obtained from the fit to data from \cite{Nakamura:2009p9} (Fig.~\ref{fig:fitnakamura}a) using $\Delta \psi=\Delta p - (k_BT/e)\log(H_{\rm int}/H_{\rm ext})$, and assuming that the maximum torque ($2500$ pN.nm) is reached at $\Delta p=-150$ mV. The apparent linear dependence of $\Delta\psi$ on external pH is consistent with previous data \cite{Lo:2007p43,Minamino:2003p532}.
\label{S1}
}
\end{figure}

Fig.~\ref{fig:fitnakamura}a shows fits of {TSRs} measured under various pH conditions \cite{Nakamura:2009p9}. The electric potential $\Delta \psi$ was not measured and so was used as a fitting parameter for each set of pH conditions.
{Our fit indicates} that $|\Delta\psi|$ increases with pH (Fig.~\ref{S1}), consistently with previous measurements \cite{Lo:2007p43,Minamino:2003p532}.

Electrorotation experiments
 \cite{Berg:1993p329,Berry:1999p403} have been used to apply an external torque on the load via an oscillating field. When the motor is driven backwards (upper-left quadrant of the {TSR}), the internal torque is approximately equal to its stall value up to speeds of $-100$ Hz \cite{Berry:1999p403}. When the motor is driven to speeds larger then the maximum operating speed (lower-right quadrant of theTSR), the motor resists rotation, resulting in a negative internal torque. In this regime, the slope of the {TSR} remains approximately the same as for positive torques beyond the knee \cite{Berg:1993p329}. Our model agrees with measurements in both regimes. The absence of a barrier to backward rotation follows from the reversibility of proton translocation. For negative torques, the model predicts an inflection of the {TSR}, as seen for the red curve in Fig.~\ref{fig:fitnakamura}a.
 
Within our model, the proton flux is limited by the loading of the external and internal gates and by the barrier crossing. This limitation on flux accounts for the knee of the {TSR}.
In Fig.~\ref{fig:fitnakamura}a, the position of the knee strongly depends on pH values: as the internal proton concentration increases, the internal gate gets saturated, preventing protons from translocating inwards, and thus limiting the stretching of stator springs and the applied torque.

\begin{figure}
\includegraphics[width=\linewidth]{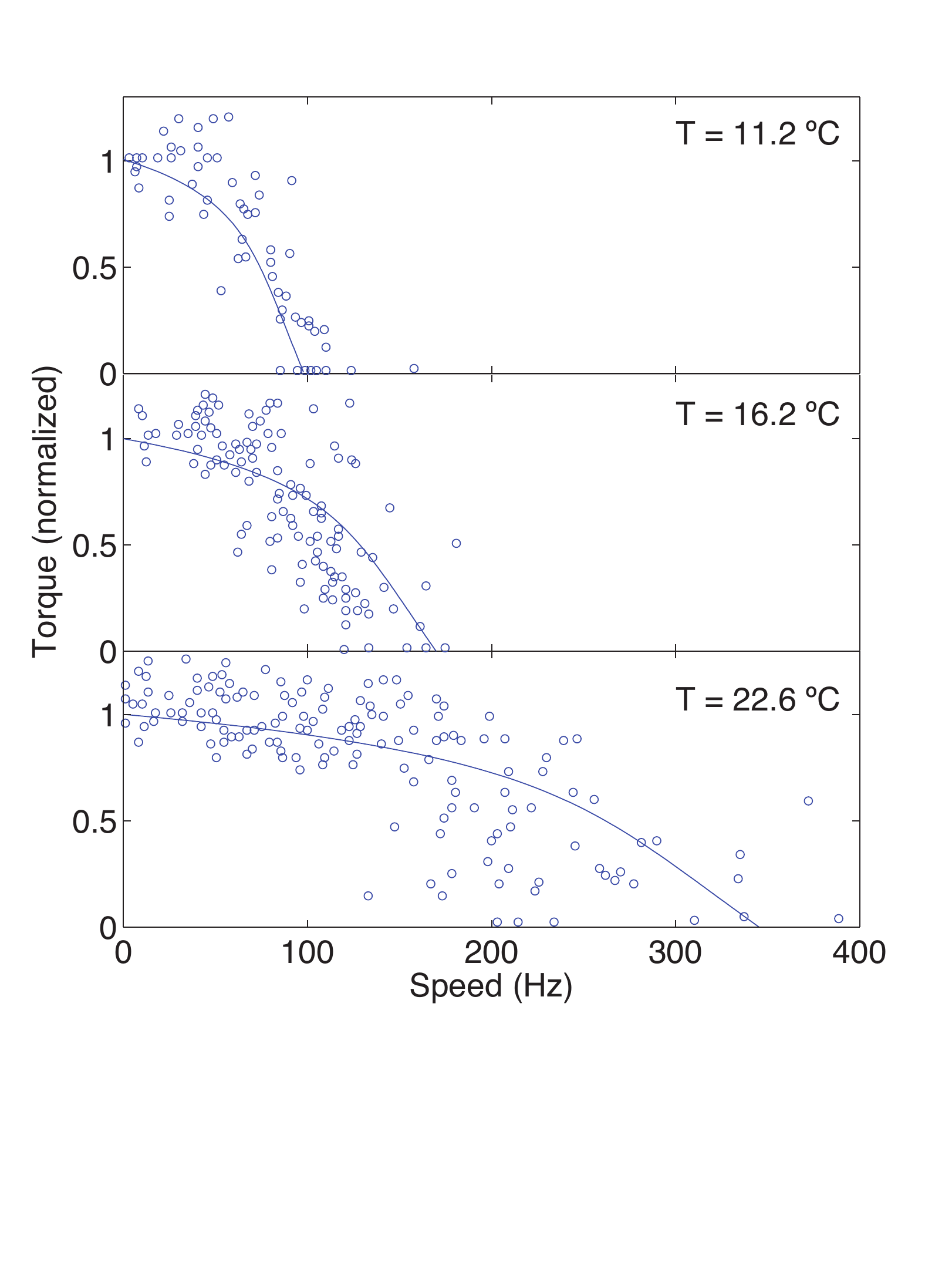}
\caption{The torque-speed relationship at various temperatures. Comparison between data from \cite{Berg:1993p329} and model prediction (solid curves) with the same parameter values as in Fig.~\ref{fig:fitnakamura}b (no additional fitting parameters). Note that the temperature of the top panel ($11.2^{o}$C) is below the range used to fit parameters (15.8-$22.7^{o}$C).
\label{S2}
}
\end{figure}

Our model also accounts quantitatively for measurements of the {TSR} at different temperatures (Fig.~\ref{fig:fitnakamura}b) \cite{Chen:2000p279}, as well as for measurements with different numbers of stators (Inset of Fig.~\ref{fig:fitnakamura}b) \cite{Ryu:2000p204}. As a test, we compared the model 
to earlier measurements at different temperatures \cite{Berg:1993p329} (data reported in  \cite{Chen:2000p279}), with no additional fitting parameters, and found excellent agreement (Fig.~\ref{S2}).

\begin{figure}
\includegraphics[width=\linewidth]{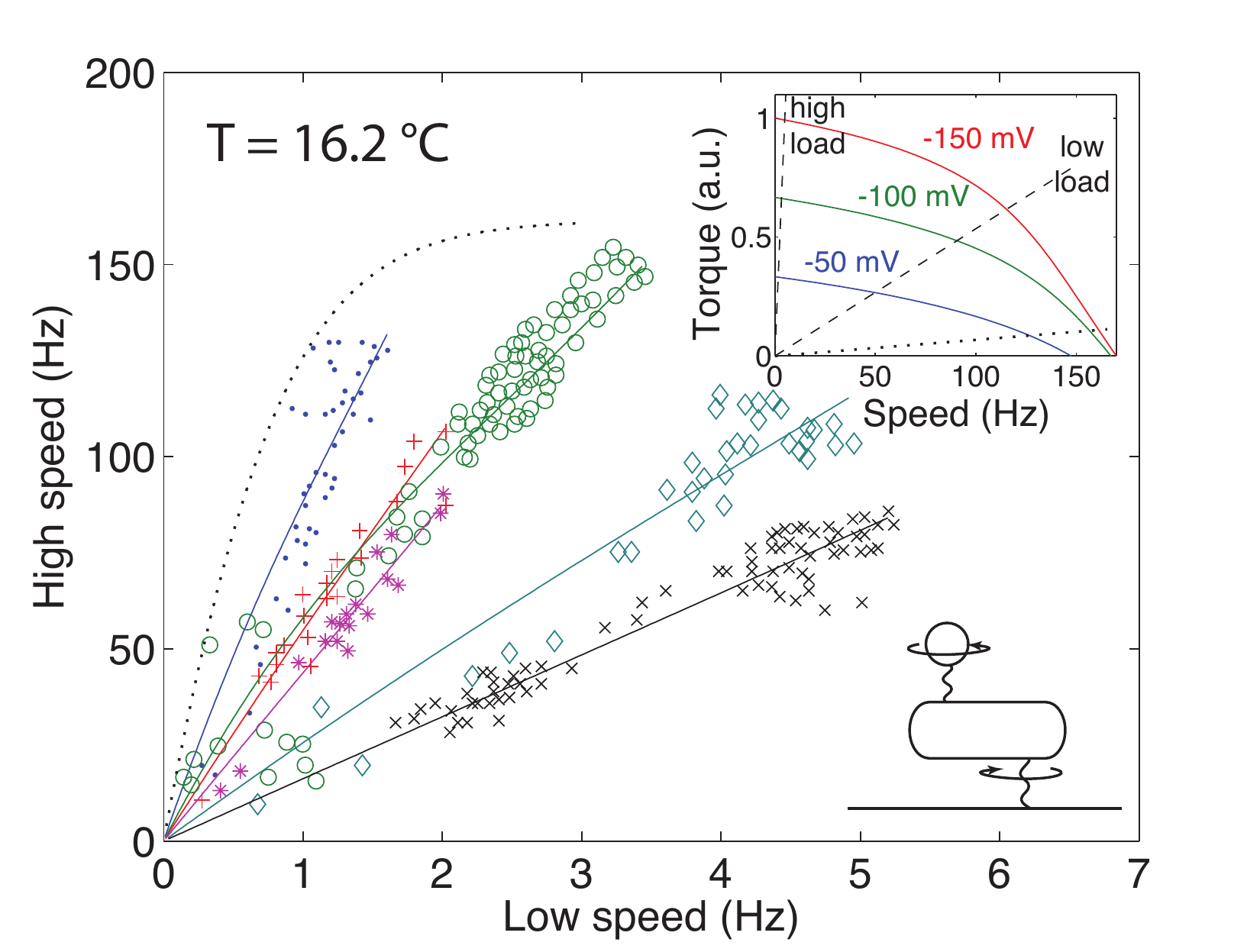}
\caption{Relationship between the high-speed (low-load) and low-speed (high-load) regimes of the {\em E. coli} flagellar motor, as the PMF varies from $-150$ mV to 0, at temperature $T=16.2\degree$C. Symbols: Experimental data from individual cells \cite{Gabel:2003p247}. Solid curves: model fits with the same parameters as in Fig.~\ref{fig:fitnakamura}. Dotted curve: the model predicts loss of proportionality at very low loads. Lower inset: schematic of the experiment \cite{Gabel:2003p247}. Upper inset: model {TSR} for PMFs of $-150$, $-100$, and $-50$ mV; the dashed lines show the low and high load lines of the cell represented by $\lozenge$ in the main panel, and the dotted line is the load line corresponding to the dotted curve in the main panel.
\label{fig:gabelfit16}
}
\end{figure}

\begin{figure}
\includegraphics[width=\linewidth]{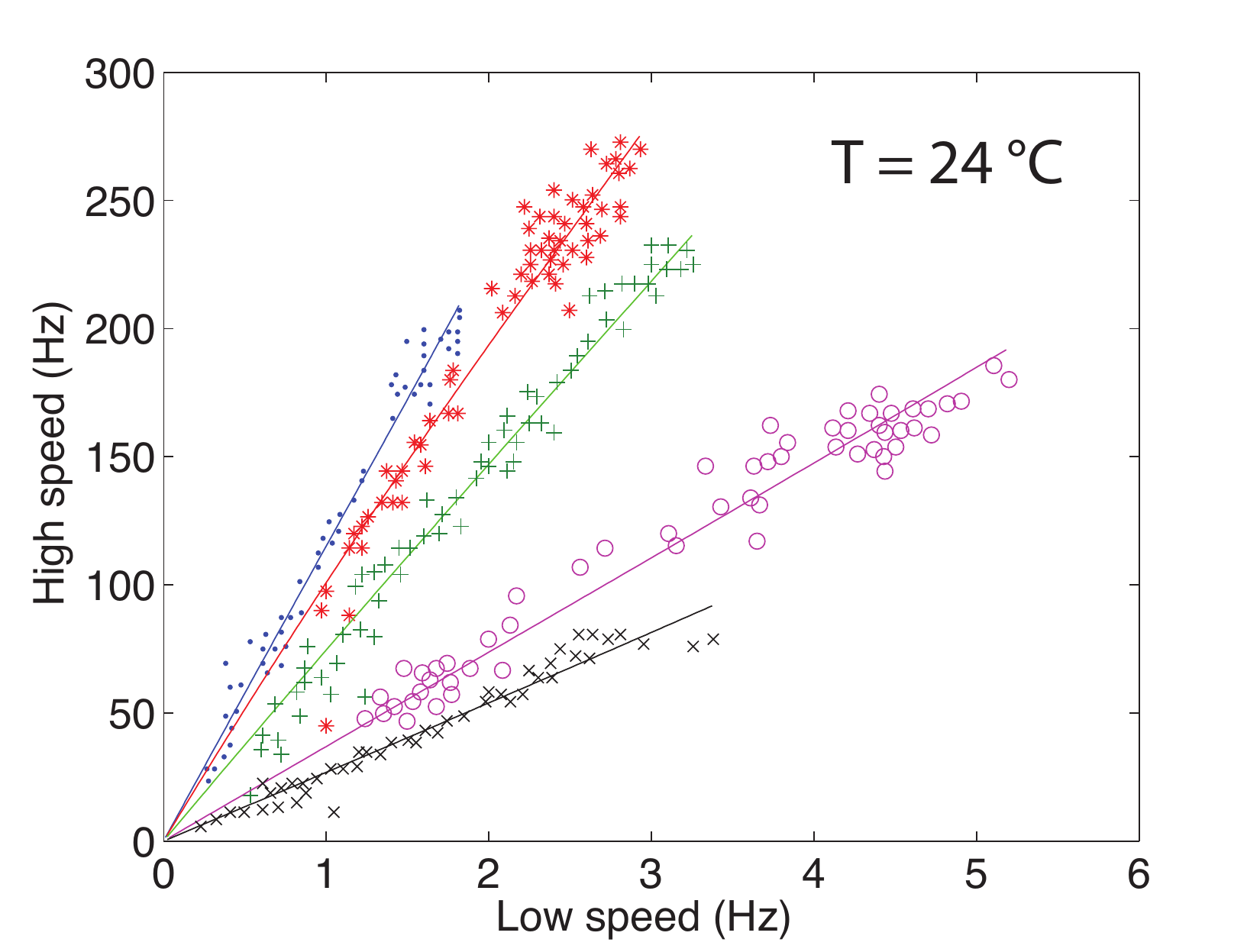}
\caption{Relationship between the high speed (low drag) and low speed (high drag) regimes of the bacterial flagellar motor, as the proton motive force varies from $-150$ to 0 mV, at temperature $T=24^{o}$C. The experimental data from individual cells \cite{Gabel:2003p247} are represented by symbols. Our fits with no proton cooperativity ($n=1$) are the solid curves.
\label{S5}
}
\end{figure}

Experiments by Gabel and Berg (Fig.~\ref{fig:gabelfit16}) \cite{Gabel:2003p247} have been interpreted to imply that the rotation speed is proportional to PMF, even at high speeds beyond the knee of the {TSR}. Our model predicts that speed is proportional to PMF at low speeds, in the plateau region of the {TSR}. However at high speeds, the torque is limited by the proton flux, and therefore both torque and speed grow sublinearly with PMF. Nevertheless, our model (with $n=1$) is fully consistent with the measurements reported in \cite{Gabel:2003p247}. 
Experimentally, speeds were simultaneously recorded for two motors of the same cell, one rotating the cell itself (high load) and the other rotating a small polystyrene bead (low load), as shown schematically in the lower inset of Fig.~\ref{fig:gabelfit16}. These two loads correspond to the two dashed lines in the upper inset of Fig.~\ref{fig:gabelfit16}. As cells were de-energized by the introduction of a respiratory poison, the PMF $\Delta p$ regressed from $-150$ mV to $0$, and the motors slowed, with the two speeds approximately proportional to each other, even at low temperature where the low-load, high-speed motor was in the kinetically-limited regime. We fitted the data for each cell, using $N/\nu_{\rm cell}$ and $N/\nu_{\rm bead}$ as free parameters, and assuming that during de-energization the electric and chemical parts of the PMF regressed in fixed ratio to each other.
The model fits are consistent with the data in the considered parameter regime, both at $24\degree$C (Fig.~\ref{S5}) and at $16.2\degree$C (Fig.~\ref{fig:gabelfit16}). However, a further reduction of the low load would be predicted to lead to a strong deviation from proportionality (dotted line in Fig.~\ref{fig:gabelfit16}).

\begin{figure}
\includegraphics[width=\linewidth]{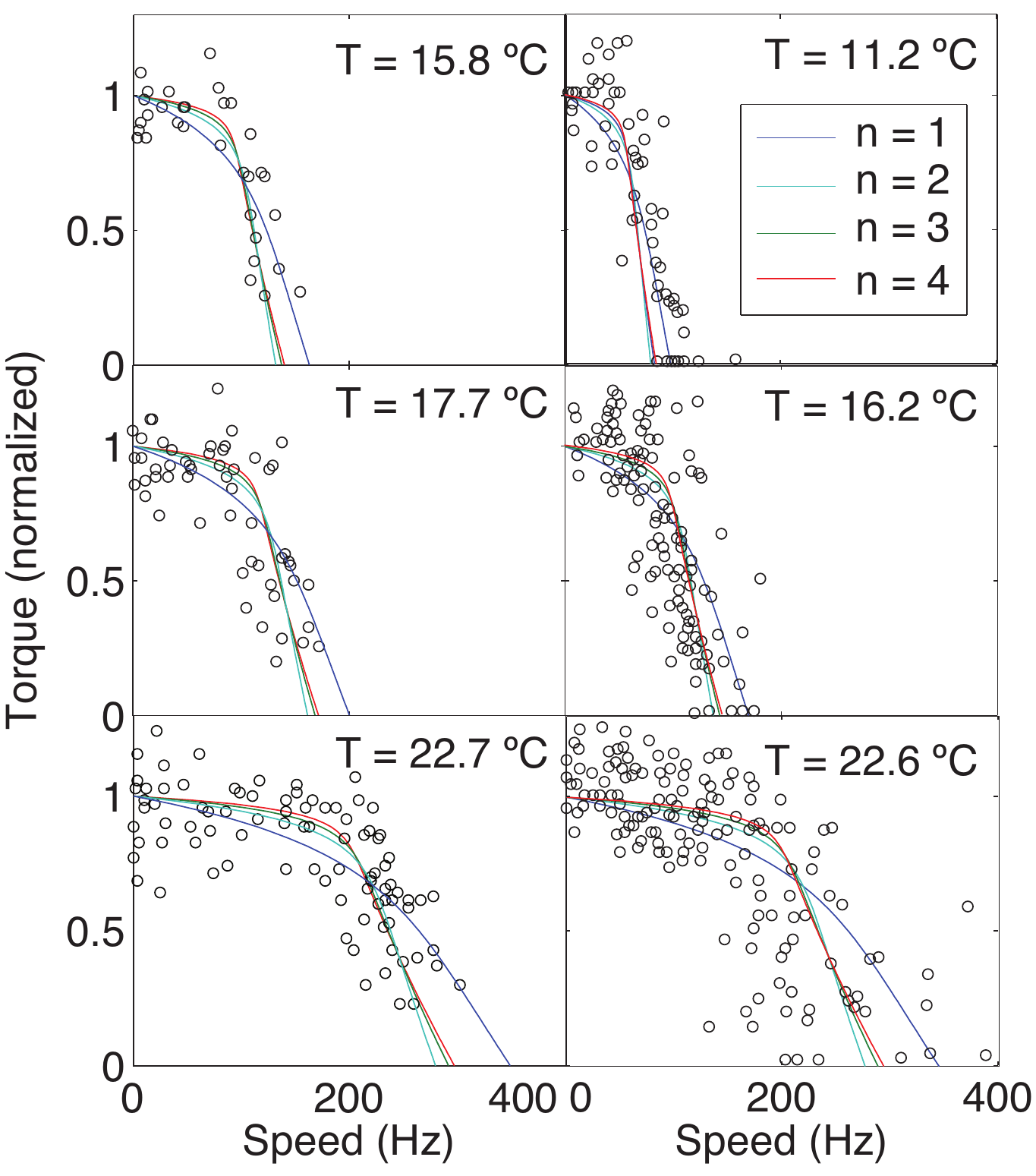}
\caption{The torque-speed relationship with proton cooperativity $n=1,\ldots,4$. Left panels: model fit to data from \cite{Chen:2000p279}. Right panels: comparison between model prediction and electrorotation data from \cite{Berg:1993p329} with the same parameter values as in the left panels.
\label{S3}
}
\end{figure}

\begin{figure}
\includegraphics[width=\linewidth]{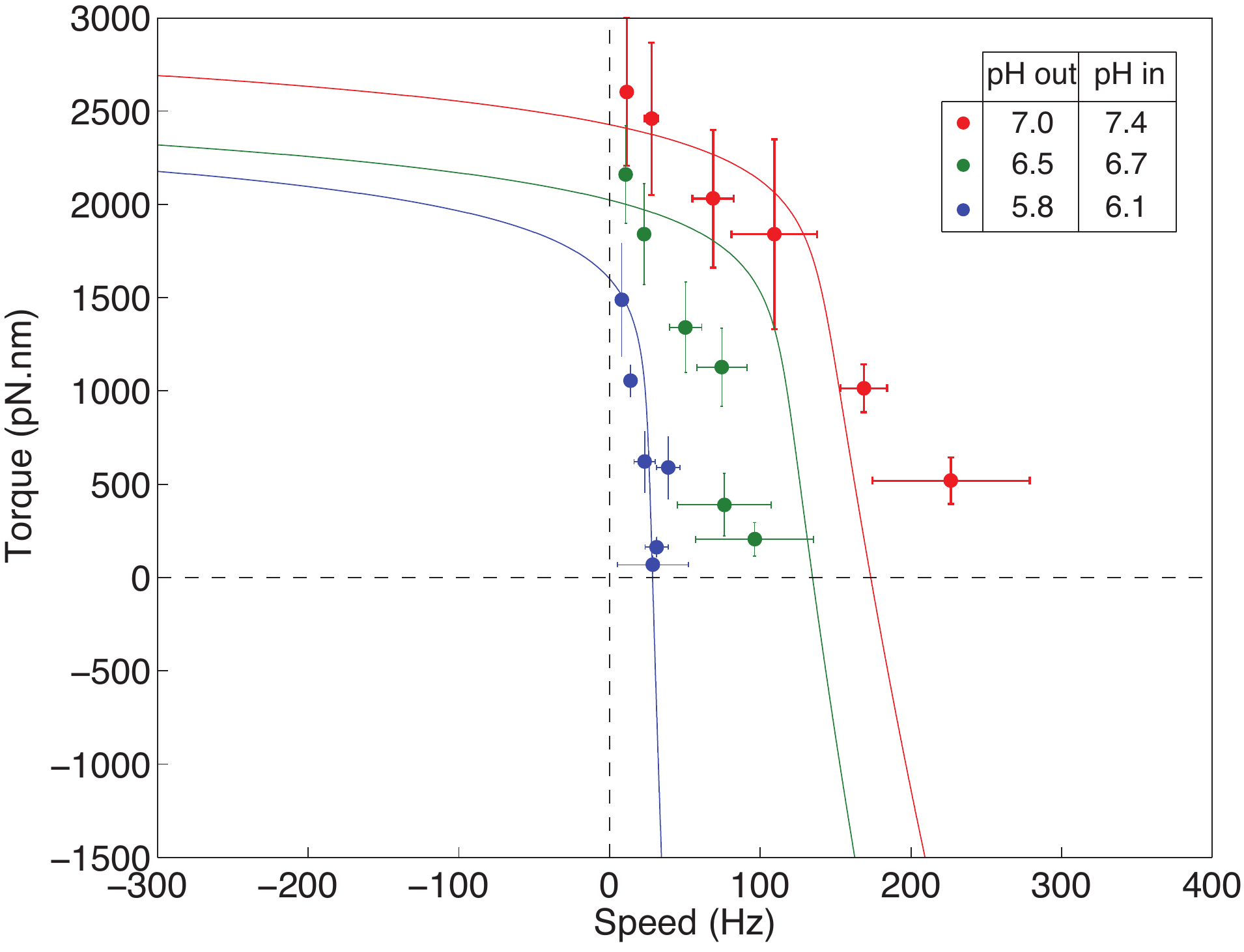}
\caption{Model fits of the torque-speed relationships from \cite{Nakamura:2009p9} with proton cooperativity $n=2$.
\label{S4}
}
\end{figure}

\begin{figure}
\includegraphics[width=\linewidth]{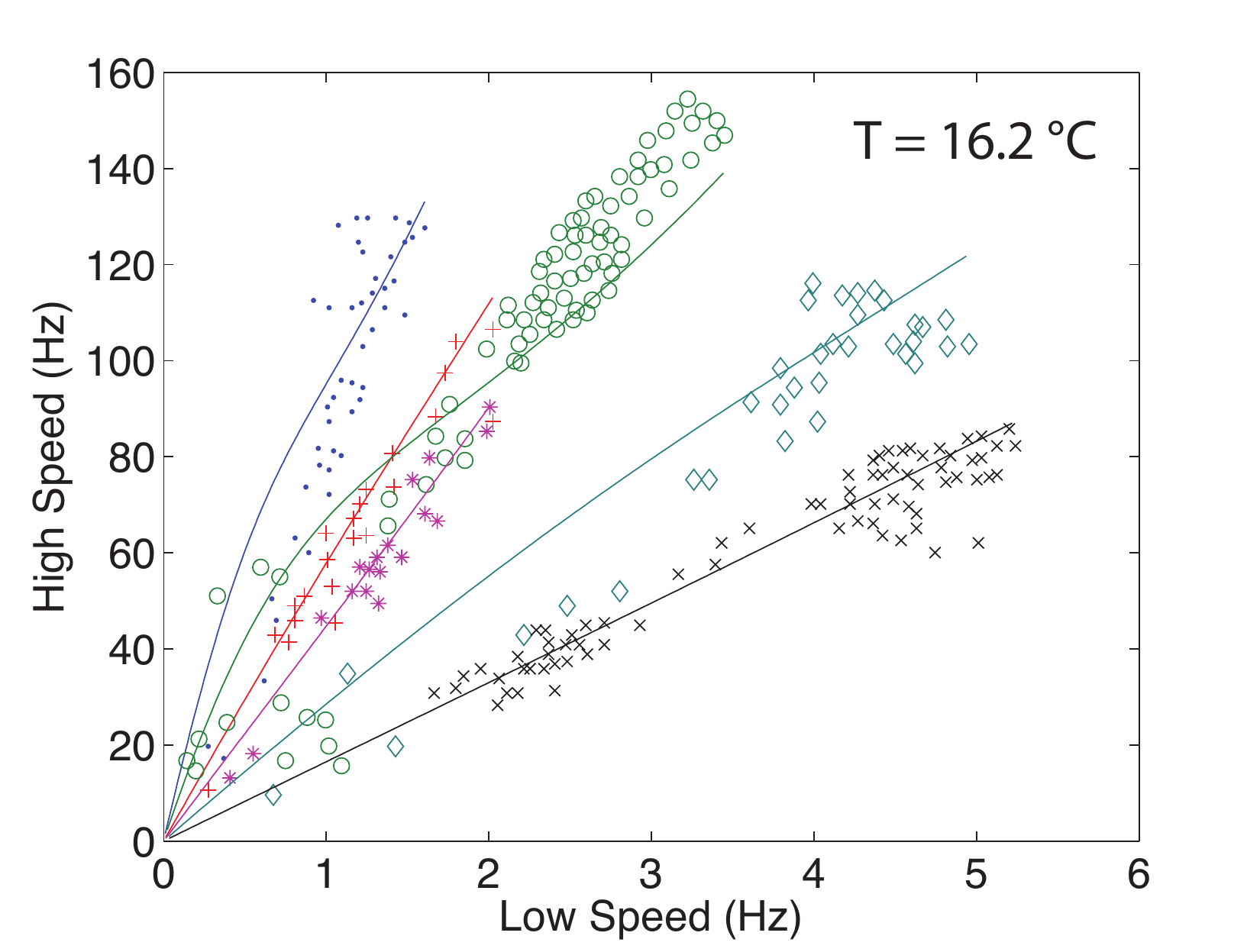}
\caption{The same high speed versus low speed data as Fig.~\ref{fig:gabelfit16}, but model fits (solid curves) obtained with proton cooperativity $n=2$.
\label{S6}
}
\end{figure}

Adding cooperativity ($n>1$) to the model still captures the general shape of the {TSR}, as well as its temperature dependence (Fig.~\ref{S3}), but agrees poorly with the observed pH dependence (Fig.~\ref{S4}), {as it implies a stronger dependence of proton current on ion concentrations}. The model with $n>1$ is consistent with Gabel and Berg's data for most cells (Fig.~\ref{S6}), although it breaks down for the two cells with the largest high speeds.

\begin{figure}
\includegraphics[width=\linewidth]{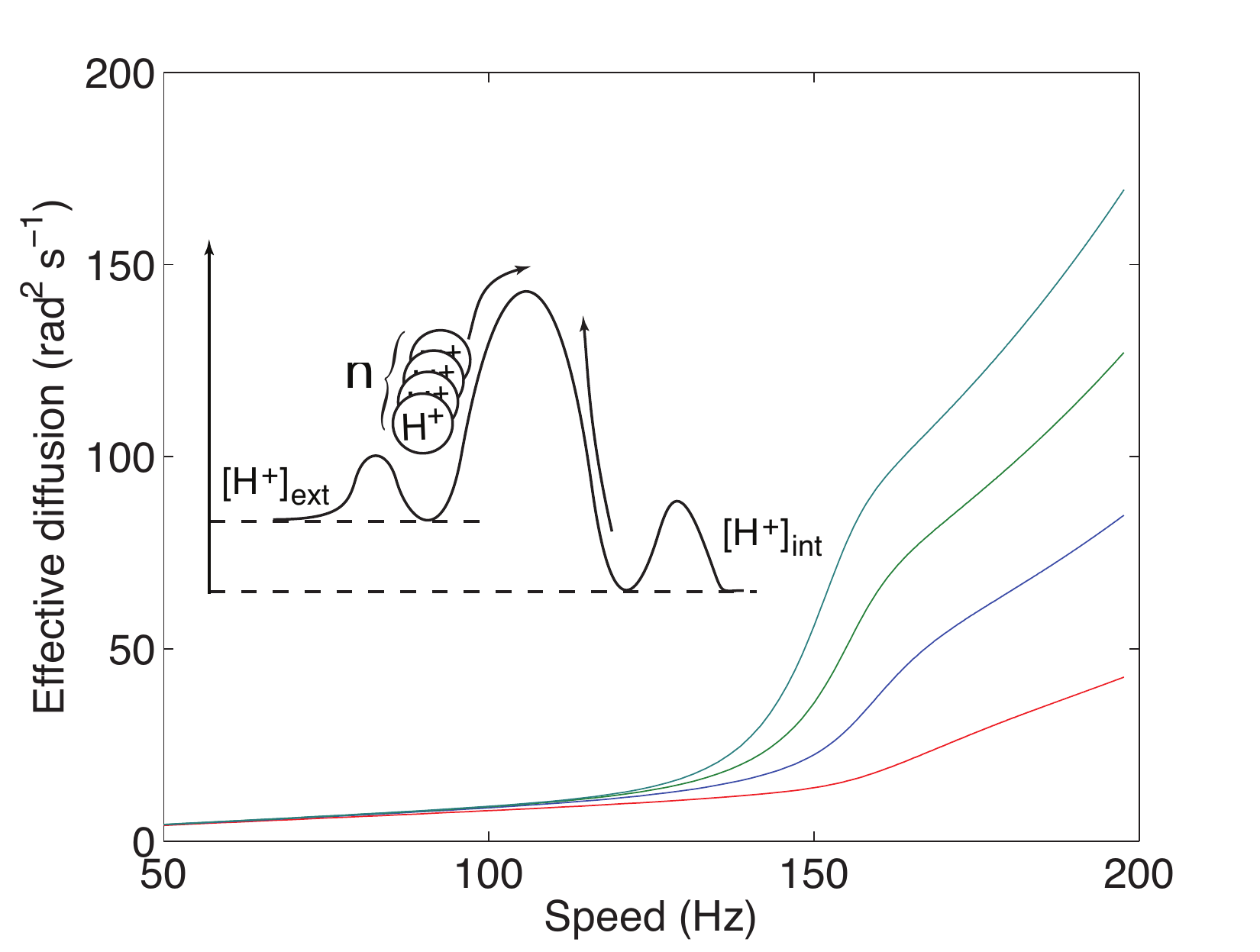}
\caption{Effective diffusion as a function of rotation speed for a single stator ($N=1$), with different ion translocation cooperativities $n=1,\ldots,4$ (bottom to top). Inset: schematic of cooperative ion translocation.
\label{fig:diffusion}
}
\end{figure}

Although $n=1$ appears to best explain the data, measurements of average rotation speeds do not allow us to discriminate with certainty the proton cooperativity. In contrast, our model predicts that diffusion of the rotor angle at low loads should depend very strongly on proton cooperativity.
The discrete nature of proton translocations implies the existence of shot noise $\xi^{\tau}_i(t)$ in the stretching of the protein spring at each stator $i$. 
We can approximate the shot noise as Gaussian white noise: $\langle \xi^{\tau}_i(t)\xi^{\tau}_i(t')\rangle=2D_{\rm shot}\delta(t-t')$, with $D_{\rm shot}= (1/2)(J_{\rm in}+J_{\rm out})\,n\delta\theta^2$ (this approximation is valid for times much larger than  $J_{\rm in}^{-1}$ and $J_{\rm out}^{-1}$).
Solving Eqs.~(\ref{eq:1},\ref{eq:2}) including shot noise $\xi^{\tau}_i(t)$ and thermal noise $\xi(t)$, we find an exact expression for the effective diffusion coefficient of the rotor angle:
\begin{equation}
\begin{split}
D_{\rm eff}&:=\lim_{t\to\infty}\frac{1}{2t}\left(\langle \theta(t)^2 \rangle-\langle \theta(t)\rangle^2\right)\\
&=\frac{k_BT}{\nu}{\left(1-\frac{N\mu}{\nu+N\mu}\right)}^2+\frac{D_{\rm shot}}{N}{\left(\frac{N\mu}{\nu+N\mu}\right)}^2,
\end{split}\label{eq:Deff}
\end{equation}
where $N$ is the number of stators and $\mu(\tau)=-(d\Omega/d\tau)^{-1}$ is minus the local slope of the {TSR}.
Fig.~\ref{fig:diffusion} shows the effective diffusion coefficient as a function of motor speed for different proton cooperativities. Parameters were chosen so that the speed at zero torque is 200 Hz.

At high loads ($\nu\gg\mu$), diffusion is entirely due to thermal noise: $D_{\rm eff}\approx k_BT/\nu$.
However, at  low loads ($\nu\ll\mu$), diffusion is dominated by shot noise: $D_{\rm eff}\approx D_{\rm shot}/N$. In fact, the thermal noise is completely suppressed in the low-load limit: {\em e.g.}, a small thermally induced backward jump in rotor angle causes the stretching of all springs, which then rapidly pull the rotor forward, thus canceling the jump. Notice that in the low-load limit, the shot-noise contribution is inversely proportional to the number of stators. Intuitively, a small jump in the angular stretch $\Delta\theta_i$ of one protein spring ultimately only causes the rotor to move $\Delta\theta_i/N$ because the rotor is equally coupled to all $N$ stators. The variance per jump is therefore $(\Delta\theta_i/N)^2$ and with $N$ independent stators, the resulting diffusion scales as $1/N$.

Eq.~\eqref{eq:Deff} could be used to infer $n$ experimentally. In the low-load limit, we have $J_{\rm in}\gg J_{\rm out}$, and therefore $n\approx 2D_{\rm eff} / (\omega \delta \theta)$. $\delta\theta$ can be determined by measuring the torque per stator and the PMF at stall: $\delta\theta=-e\Delta p/\tau$.

Our analysis of rotor diffusion suggests a novel experimental test to investigate the cooperativity of proton translocation. Some rotational diffusion measurements have already been made \cite{Samuel:1995p234,Samuel:1996p226}, but not in the regime of very low load, where shot noise is expected to dominate.
Although we have derived the expression for diffusion in the specific framework of our minimal spring model, the same approach is generalizable to more detailed models of the bacterial flagellar motor.

We thank Avigdor Eldar and Anirvan Sengupta for helpful suggestions.
T.M. was supported by a Human Frontier Science
Program fellowship, and N.S.W.  by National Institutes
of Health Grant R01 GM082938.

%

%


\bibliographystyle{apsrmp}

\bibliography{bfmnourl,bfm2}

\end{document}